\documentclass[aps,prd,twocolumn,nofootinbib]{revtex4-2}

\usepackage{amsmath,amssymb,graphicx,bm}
\usepackage{xcolor,hyperref,booktabs}
\usepackage{array,multirow}
\usepackage[separate-uncertainty=true]{siunitx}

\begin{document}

\title{Gravitational-Wave Inference For Noise PSD Jumps Across Data Gaps}

\author{Jian-Ming Yan}
\email{yanjm@ucas.ac.cn}
\affiliation{School of Fundamental Physics and Mathematical Sciences, Hangzhou Institute for Advanced Study, University of Chinese Academy of Sciences, Hangzhou 310024, China}

\author{Zong-Kuan Guo}
\email{guozk@itp.ac.cn}
\affiliation{School of Fundamental Physics and Mathematical Sciences, Hangzhou Institute for Advanced Study, University of Chinese    Academy of Sciences, Hangzhou 310024, China}
\affiliation{Institute of Theoretical Physics, Chinese Academy of Sciences (CAS), Beijing 100190, China}
\affiliation{University of Chinese Academy of Sciences (UCAS), Beijing 100049, China}

\date{\today}

\begin{abstract}
Long-duration gravitational-wave (GW) inference inevitably encounters data gaps, which are often accompanied by abrupt, non-stationary changes in the detector noise power spectral density (PSD). Conventional practice typically analyzes the pre-gap and post-gap data separately, causing avoidable information loss and potentially degrading parameter inference. We build on Bayesian gap augmentation in the Wilson--Daubechies--Meyer (WDM) time--frequency domain and focus on a practically important failure mode: short gaps with large PSD amplitude jumps. In this regime, the local diagonal (Whittle-like) approximation that underpins the WDM likelihood can no longer be relied upon in the time direction near the gap edges, which may lead to biased and/or less robust inference unless the local diagonal validity is restored without resorting to full segmentation. We propose a two-stage, validity-driven remedy. First, frequency-domain prior-predictive prewhitening (PW) incorporates endpoint-informed noise information to mitigate the dominant mismatch responsible for poor diagonal behavior, enabling a more favorable WDM representation while retaining computational tractability. Second, when a time-direction smoothness criterion still fails, we apply adaptive gap expanding (AGE), selectively enlarging only the minimal neighborhood around the gap boundaries needed to restore local diagonal validity. Toy-model simulations with chimeric PSD transitions show that WDM+PW+AGE yields substantially tighter and more accurate posteriors than single-sided pre-gap/post-gap analyses, while maintaining computational efficiency suitable for next-generation missions.
\end{abstract}

\maketitle

\section{Introduction}
\label{sec:intro}

Long-duration gravitational-wave observations with next-generation detectors
will be limited not only by sensitivity but also by data realism. Since the first
direct detection by the ground-based LIGO--Virgo network~\cite{LIGO2016} and the
subsequent observations that established GW astronomy~\cite{Abbott2018,LIGO2019},
the field has extended its reach from the $\sim10$--$10^4\,\mathrm{Hz}$
ground-based band into the millihertz regime. This band will be monitored by a
fleet of planned missions: together with LISA~\cite{LISA2017}, the Chinese-led
Taiji~\cite{Taiji2017,Taiji2021} and TianQin~\cite{TianQin2016,TianQin2021}
programs are scheduled for launch in the 2030s, with complementary orbital
configurations and sensitivity windows; a jointly operating LISA--Taiji network
would substantially improve sky localization and parameter
reconstruction~\cite{LISATaiji2020}. Key technologies have already been validated
in orbit by the Taiji~\cite{Taiji2021FirstStep} and TianQin~\cite{TianQin2021}
pathfinders. These missions target broadly similar source populations, including
massive black-hole binaries~\cite{Arun2009,Klein2016,TianQinMBH2019},
extreme-mass-ratio inspirals~\cite{Gair2010,Babak2017}, and millions of galactic
binaries~\cite{Timpano2006,Cornish2017}. Third-generation (3G) ground-based detectors
such as the Einstein Telescope and Cosmic Explorer~\cite{ET2020,CE2023} will extend
the accessible band below $10\,\mathrm{Hz}$, opening continuous-wave and multiband
studies~\cite{Sesana2016}. These developments have motivated community benchmarks
such as the LISA Data Challenges~\cite{Baghi2022LDC} and the Taiji Data
Challenges~\cite{TaijiTDC1,TaijiTDC2}.

A defining feature of these missions is that signals persist from weeks to years,
while detector noise is intrinsically time-varying. The stationary-noise assumptions
underlying classical GW inference~\cite{Cutler1998,Whittle1954} therefore become
progressively less reliable as non-stationarities evolve on timescales comparable to
the analysis baseline. Instrumental effects include laser frequency and intensity
fluctuations~\cite{Tinto2002,Tinto2024}, attitude-control maneuvers and antenna
re-pointing~\cite{Burke2025TDI}, and cryogenic thermal cycling
in 3G instruments and KAGRA~\cite{ET2020,KAGRA2020}. In addition, time-varying
astrophysical foregrounds---such as the annually modulated galactic-binary confusion
noise produced by LISA's orbital motion---act as effectively non-stationary
components~\cite{Digman2022,Cornish2003}. Data-processing steps can introduce further
artifacts~\cite{Bayle2023}, and the orbital response and armlength variations of
LISA~\cite{Cornish2003,Larson2000,Babak2021SNR} complicate noise modeling beyond a
simple PSD-in-time description. The same difficulties are generic to the Chinese
millihertz missions: Taiji and TianQin both have unequal, continuously flexing
armlengths that require second-generation time-delay interferometry
(TDI)~\cite{TaijiTDC1,TianQinTDI2021}, and recent Taiji challenge studies explicitly
identify glitches, instrumental drifts, cyclostationary astrophysical foregrounds,
and gaps as unavoidable in-orbit conditions, noting in particular that non-stationarity
violates the assumptions of the conventional Whittle likelihood and greatly increases
the cost of Bayesian inference~\cite{TaijiTDC2}. Dedicated Bayesian treatments of
non-Gaussian transients and instrument glitches, such as the BayesWave algorithm,
have been developed for ground-based data and are directly relevant to this
setting~\cite{Cornish2015}. Global-fit
pipelines~\cite{Katz2024,Deng2025,Strub2024,Littenberg2023} therefore require
likelihoods that are simultaneously tractable and consistent with realistic
non-stationarities.

Compounding non-stationarity, data gaps---intervals of interrupted
measurements---are unavoidable. For LISA the duty cycle is expected to remain
appreciably below unity, so that a non-negligible fraction of the mission is
lost to gaps. Comparable assessments for Taiji point to a similar situation,
with gaps arising from both scheduled maintenance and maneuvers (such as
antenna re-pointing and adjustments of the frequency plan and pointing scheme)
as well as from unscheduled in-orbit disruptions~\cite{TaijiTDC2,Pearson2025}.
Gap handling is further complicated by TDI: Burke et al.~\cite{Burke2025TDI}
showed that even a single missing sample can expand into a much longer
effective gap in second-generation TDI variables, causing a non-negligible
loss of usable data. Non-stationary noise and gaps together create pathologies
for standard approaches. First, sharp gap edges induce spectral
leakage~\cite{Baghi2019}: the Fourier coefficients of the gapped data are no
longer uncorrelated, so the independence assumptions behind Fourier-domain
likelihoods break down. Tapering the data or the templates suppresses such
leakage, at the price of discarding part of the signal power~\cite{Ajith2010}.
In this setting the Whittle likelihood~\cite{Whittle1954},
widely used in GW parameter estimation, where posteriors are commonly sampled with
Markov chain Monte Carlo (MCMC) and trans-dimensional model choices with reversible-jump
Monte Carlo~\cite{Veitch2015,Green1995}, can yield
biased---and, under PSD misspecification, statistically
inconsistent---estimates, particularly for strongly colored
noise~\cite{Burke2025,Castelli2024}. Second, non-stationarity destroys the
diagonal structure of the covariance in the Fourier domain, making exact
covariance treatments computationally prohibitive at the scale of these data
sets and motivating approximate representations.

This paper starts from two closely related lines of work. First, gaps have been addressed by Bayesian data augmentation: Baghi et al.~\cite{Baghi2019} treat the missing samples as auxiliary variables and impute them from their conditional posterior within a blocked Gibbs sampler, so that inference
proceeds on the completed data set even though the covariance of the observed samples
alone is neither Toeplitz nor diagonal in the Fourier domain. Second, and most relevant
here, Pearson and Cornish~\cite{Pearson2025} extended this augmentation to the
time--frequency domain. Working in the Wilson--Daubechies--Meyer
basis~\cite{Necula2012}, an orthonormal time--frequency decomposition suited to locally
stationary processes, the wavelet-domain noise covariance is approximately diagonal
provided the PSD varies slowly across each wavelet pixel~\cite{Cornish2025}; this
near-diagonal structure makes Gaussian conditional imputation feasible at a cost
governed by a local neighborhood rather than by the full data set.

These are not the only routes to gap-robust inference. Complementary approaches include time-domain windowing and inpainting for TianQin~\cite{TianQinGap2025}, deep-learning-based imputation for non-stationary noise~\cite{Mao2025,Xu2024}, and TDI-based methods robust to missing samples~\cite{Houba2025}. Building on the framework just outlined—Bayesian gap augmentation in the WDM basis with a chimeric PSD transition—this work addresses a key failure case: large PSD amplitude jumps across a gap. Such jumps can break the time-direction smoothness assumption, reducing the reliability of the Whittle-like diagonal approximation particularly in tiles overlapping the transition. Simply ignoring this approximation loss may bias inference, while full segmentation discards substantial information. This motivates a mechanism that restores local diagonal validity without the full cost of segmentation.

To this end, building on the WDM augmentation framework reviewed above, we
propose a two-stage strategy combining frequency-domain prior-predictive
prewhitening with adaptive gap expanding. PW uses prior-predictive
PSD information from data adjacent to each gap to rebalance the WDM
smoothness requirements, enabling finer time resolution near gap edges under a
fixed budget. When the time-direction criterion is still not satisfied, AGE
selectively removes only a minimal neighborhood around the gap boundaries to
restore local diagonal validity. Toy-model experiments with chimeric noise and
linear-chirp injections indicate that the combined strategy produces more
accurate posteriors than single-sided analyses while discarding far less data
than a full segmentation fallback.

The remainder of this paper is organized as follows.
Sec.~\ref{sec:prewhiten_and_gap} first summarizes the WDM gap-augmentation framework
underlying our method and explains why it fails in the short-gap/large-jump regime, and
then presents our two-stage remedy: frequency-domain prior whitening and adaptive gap
expanding. Sec.~\ref{sec:test} introduces the toy model used to isolate and stress-test
this regime, and reports the numerical results. We conclude in
Sec.~\ref{sec:conclusion}.

\section{Handling Short Gaps with Large PSD Jumps in GW Inference}
\label{sec:prewhiten_and_gap}

This section presents our method. We review the WDM gap-augmentation framework of
Pearson and Cornish~\cite{Pearson2025}—its wavelet-domain near-diagonal covariance and the
chimeric PSD transition across a gap—and show why it fails for short gaps with large PSD jumps.
We develop a two-stage remedy: frequency-domain prior-predictive prewhitening (Stage~1),
followed by adaptive gap expanding (Stage~2), which restores local diagonal validity (with a
minimal guard band) only if the time-direction criterion remains violated after Stage~1.

\subsection{Wavelet-domain data augmentation with chimeric transitions}

Pearson and Cornish~\cite{Pearson2025} extend gap augmentation to the
time--frequency domain. Conceptually, this is the same augmentation idea introduced by
Baghi et al.~\cite{Baghi2019}: the unobserved data are imputed from their conditional posterior
rather than being discarded. Data augmentation is itself a classical device in Bayesian
statistics~\cite{Tanner1987,Albert1993}, and the conditional imputation step is naturally
embedded in a Gibbs sampler~\cite{Geman1984,Gelfand1990}. The difference here is that the
imputation is carried out on wavelet coefficients instead of time samples. The central idea is to work in the
WDM wavelet basis, where a non-stationary but locally
stationary noise process can be approximated by a near-diagonal covariance. It is this
near-diagonal structure that makes Gaussian conditional imputation computationally feasible,
since the conditioning can then be restricted to a local neighborhood instead of requiring the
inversion of a dense covariance matrix.

\paragraph{WDM representation (basic formulas).}
Let the (zero-mean) time-series data be $d[k]$ with $k=0,\dots,N-1$.
In WDM, the data are projected onto an orthonormal (or near-orthonormal) set of
time--frequency basis functions $\{\psi_{nm}[k]\}$, yielding wavelet coefficients
\begin{equation}
w_{nm} \equiv \langle d,\psi_{nm}\rangle
= \sum_{k=0}^{N-1} d[k]\;\psi_{nm}[k],
\label{eq:wdm_coeff_def}
\end{equation}
where $n=0,\dots,N_t-1$ indexes time pixels and $m=0,\dots,N_f-1$ indexes frequency layers.
Stacking all coefficients produces a vector $\mathbf{w}$ of length $N=N_tN_f$.
For Gaussian noise, the wavelet-domain covariance
\begin{equation}
\Lambda_{(nm)(n'm')} \equiv \mathbb{E}\!\left[w_{nm}\,w_{n'm'}\right]
\label{eq:wdm_cov_def}
\end{equation}
fully determines the likelihood and the conditional distributions used in augmentation.

\paragraph{Why wavelets help.}
For a locally stationary noise process, the time-dependent PSD $S(f,t)$
is approximately constant within each wavelet pixel.
In this regime, wavelet coefficients decorrelate across different time--frequency pixels,
and the covariance in Eq.~\eqref{eq:wdm_cov_def} can be approximated as
\begin{equation}
\Lambda_{(nm)(n'm')} \approx S(f_m,t_n)\,\delta_{nn'}\delta_{mm'},
\label{eq:wdm_diagonal_simplified}
\end{equation}
where $S(f_m,t_n)$ is the representative PSD value in the $(n,m)$ pixel.
This diagonal (Whittle-like) approximation implies that the conditional distribution of missing
coefficients given observed ones is again Gaussian, with a mean/covariance that can be computed
using only coefficients in a local neighborhood rather than the full dense matrix
(see~\cite{Pearson2025} for the detailed conditioning strategy).

Concretely, partitioning wavelet coefficients into observed and missing subsets
$\mathbf{w}=(\mathbf{w}_o,\mathbf{w}_m)$, the conditional distribution takes the Gaussian form
\begin{align}
p(\mathbf{w}_m \mid \mathbf{w}_o, \boldsymbol{\theta})
&= \mathcal{N}\!\left(\boldsymbol{\mu}_{m|o},\,\mathbf{\Sigma}_{m|o}\right),
\label{eq:gaussian_conditional_wdm}\\
\boldsymbol{\mu}_{m|o}
&= \mathbf{h}_m(\boldsymbol{\theta})
+ \mathbf{\Sigma}_{mo}\mathbf{\Sigma}_{oo}^{-1}\left(\mathbf{w}_o-\mathbf{h}_o(\boldsymbol{\theta})\right),\\
\mathbf{\Sigma}_{m|o}
&= \mathbf{\Sigma}_{mm}-\mathbf{\Sigma}_{mo}\mathbf{\Sigma}_{oo}^{-1}\mathbf{\Sigma}_{mo}^{T},
\label{eq:gaussian_conditional_wdm_cov}
\end{align}
where $\mathbf{h}$ is the signal model mapped into the same wavelet basis.
The cost is reduced because, under the near-diagonal structure in Eq.~\eqref{eq:wdm_diagonal_simplified},
only coefficients whose wavelet basis functions overlap the missing time region contribute appreciably.

The locality is controlled by the wavelet filter length:
in the WDM construction used by Pearson and Cornish~\cite{Pearson2025},
\begin{equation}
K = 2qN_f,
\label{eq:K_def}
\end{equation}
where $q$ is a scaling parameter setting the time support relative to the pixelization
and $N_f$ is the number of frequency layers. Hence, a basis function centered near a given time pixel
has support over approximately $K$ samples in time, corresponding to an effective half-width
$\sim K/2$.
Therefore, a gap edge affects a neighborhood of coefficients whose time supports intersect the gap,
so conditional imputation can be performed using coefficients localized in time--frequency space.

\paragraph{Melding chimeric PSD transitions.} 
When the  PSD changes across a gap, a single smooth connector can still represent a relatively rapid state change within the time span covered by the relevant WDM wavelet supports. In such cases,
the local assumptions that motivate the WDM diagonal (Whittle-like) approximation
may be harder to maintain in the time--frequency tiles overlapping the transition. 

Pearson and Cornish address this by introducing a chimeric PSD model, which blends
two disjoint (pre-gap and post-gap) PSD descriptions through time using smoothly
varying transition functions: 
\begin{equation}
S^{\rm chimera}(f,t) = S_1(f,t)\,w_1(t) + S_2(f,t)\,w_2(t),
\label{eq:chimera_psd}
\end{equation}
where $S_1$ and $S_2$ represent the pre-gap and post-gap PSDs, respectively, and
$w_1(t)$, $w_2(t)$ are smooth blending functions that interpolate across a connector
region. By performing the blending over multiple WDM time pixels, the PSD change is
represented as a gradual transition over the wavelet-support region, providing a
more consistent description of the noise state change for the gap-filling procedure. 

In this framework, however, the approach can become more difficult as the gap becomes
shorter and the difference between the two noise models becomes larger, since this places
additional stress on the locally-stationary assumption underlying the diagonalization in
the wavelet domain.

\paragraph{Local validity from smoothness parameters.}
The diagonal approximation holds only when the log-derivatives of the PSD are small
on the scale of a wavelet pixel~\cite{Cornish2025}.
Cornish quantifies this using dimensionless smoothness measures
\begin{align}
s_1(f,t) &\equiv \Delta F\,\left|\partial_f \ln S(f,t)\right|,\\
\mu_1(f,t) &\equiv \Delta T\,\left|\partial_t \ln \sigma^2(t)\right|,
\label{eq:cornish_smoothness}
\end{align}
where $\Delta F$ and $\Delta T$ are the frequency and time pixel widths, and $\sigma^2(t)$ denotes
the characteristic noise variance (or equivalently an amplitude factor of the PSD) over the time pixel
used in the wavelet construction.
When both $|s_1|$ and $|\mu_1|$ remain sufficiently small (typically less than $\sim 10\%$),
the off-diagonal residuals in the wavelet-domain covariance are suppressed
to $\lesssim 1\%$, and the resulting Whittle-like (diagonal) WDM likelihood yields accurate
parameter inference~\cite{Pearson2025,Cornish2025}.

\subsection{How large PSD jumps break the diagonal approximation}
\label{subsec:wdm_failure}

Despite the progress above, a practically important failure mode remains:
\textbf{short gaps accompanied by large PSD amplitude jumps}.

In such scenarios:
\begin{itemize}
\item The gap is short (e.g., $\sim 2$ hours in a 7-day observation), so the connector region $L$ spans only a small fraction of the WDM time support;
\item The PSD amplitude ratio $(A_{\rm post}/A_{\rm pre})$ is large (e.g.\ $2$--$3$ or more);
\item Even with a smooth connector, the time-direction log-derivative $\partial_t\ln A$ becomes large and can violate the WDM diagonal-validity condition controlled by $\mu_1$.
\end{itemize}

Quantitatively, from the chimeric amplitude model
\begin{equation}
A(t)=A_{\rm pre}+(A_{\rm post}-A_{\rm pre})\,w(u(t)),
\end{equation}
with $u(t)$ varying over a transition width $L$, we have in the transition region
\begin{equation}
\begin{aligned}
\left|\partial_t \ln A\right|
&=
\left|\frac{\partial_t A}{A}\right|
=
\left|
\frac{A_{\rm post}-A_{\rm pre}}{A(t)}\cdot \frac{w'(u)}{L}
\right| \\
& \sim
\frac{|A_{\rm post}-A_{\rm pre}|}{A(t)}\cdot\frac{|w'(u)|}{L}.
\end{aligned}
\label{eq:amplitude_deriv_scaling}
\end{equation}
Therefore the time-direction smoothness parameter scales as
\begin{equation}
\begin{aligned}
\mu_1 &\equiv \Delta T\,\left|\partial_t \ln \sigma^2(t)\right|
\propto\Delta T\,\left|\partial_t \ln A(t)\right| \\
&\sim \Delta T\,\frac{|A_{\rm post}-A_{\rm pre}|}{A(t)}\cdot\frac{|w'(u)|}{L}.
\end{aligned}
\end{equation}
Since $\Delta T$ is the characteristic WDM time-pixel width (fixed once $(N_f,\Delta t)$ are chosen), a short $L$ together with a large jump can push $\mu_1$ above the diagonal-validity threshold even when $w$ is chosen to be smooth.

\paragraph{What ``resolution'' would ideally help: increase $N_t$ (smaller $\Delta T$).}
To reduce $\mu_1$, one would like to make the WDM time pixels narrower, i.e.\ to reduce $\Delta T$.
Given the WDM pixelization used here, $\Delta T$ is controlled by $N_f$ (through $\Delta T=N_f\Delta t$), so reducing $\Delta T$ corresponds to choosing smaller $N_f$.
Under the fixed total sample budget constraint $N=N_tN_f$, this is equivalent to choosing larger $N_t$.
Thus, from the viewpoint of the time failure mode, the appropriate direction is indeed $N_t\uparrow$ (or $N_f\downarrow$).

\paragraph{The paradox: time and frequency requirements conflict under fixed budget.}
However, decreasing $N_f$ does not come for free: it increases the frequency pixel width $\Delta F=1/(2N_f\Delta t)$ and hence tightens the frequency-direction smoothness requirement controlled by $s_1$~\cite{Cornish2025}. As a result, the diagonal (Whittle-like) WDM likelihood fails locally near the gap boundaries, biasing the inference unless additional measures are taken~\cite{Pearson2025}.

\paragraph{How we escape the conflict.}
The analysis above identifies the precise bottleneck: under a fixed budget $N=N_tN_f$, time resolution can only be bought at the price of frequency resolution, so the jump-induced $\mu_1$ violation cannot be removed by re-pixelizing alone.
Our remedy attacks the other side of the balance instead.
Rather than increasing $N_f$ to absorb a large spectral dynamic range, we first remove most of that dynamic range by prewhitening the data with a fixed reference spectrum built from prior-predictive endpoint information (Stage~1 below).
Because the frequency-direction requirement is set by the residual log-slope after whitening, a much coarser frequency resolution---hence a finer time resolution at fixed $N$---suffices, which in turn suppresses $\mu_1$ at the gap edges.
Only if the time-direction criterion is still violated after this rebalancing do we invoke a second, deliberately minimal measure: we discard the smallest neighborhood around the gap edges that restores local diagonal validity (Stage~2 below), instead of splitting the data set into two independent segments.

\subsection{Stage 1: frequency-domain prior-predictive prewhitening}

In the short-gap + large-PSD-jump regime, the dominant failure comes from
violations of the WDM diagonal-validity conditions near the gap edges.
To make the Whittle-like (diagonal) WDM likelihood as robust as possible
while avoiding unnecessary data loss, we adopt a two-stage strategy that
explicitly targets the reduction of the required number of WDM frequency layers $N_f$.

In the WDM framework, the wavelet-domain noise covariance is approximately diagonal
when the dynamic PSD varies slowly within each wavelet pixel.
The leading contribution to off-diagonal residuals is controlled by the
dimensionless frequency-direction smoothness parameter~\cite{Cornish2025},
\begin{equation}
s_1 \equiv
\Delta F\,\left|\partial_f \ln S(f,t)\right|,
\qquad
\Delta F=\frac{1}{2N_f\,\Delta t},
\label{eq:s1_def_rewrite}
\end{equation}
where $N_f$ is the number of WDM frequency layers and $\Delta t$ is the sampling interval.
For a prescribed tolerance $\varepsilon$, diagonal (Whittle-like) accuracy is obtained when
$|s_1|\lesssim \varepsilon$ (e.g., $\varepsilon = 0.1$ ).

Because $\Delta F\propto 1/N_f$, satisfying $|s_1|\lesssim \varepsilon$ typically
forces $N_f$ to increase whenever the PSD transition induces
unexpectedly large frequency-direction log-slopes.
Under the fixed computational budget $N=N_tN_f$, reducing $N_f$
directly improves the attainable time resolution $N_t$ and
makes it easier to satisfy the time-direction smoothness criterion.
Therefore, we seek to reduce the worst-case frequency log-slope that enters $s_1$
in the planning stage.

\subsubsection{Step 1a: Endpoint inference and construction of a tight noise prior}
Let $\boldsymbol{\theta}$ denote the (possibly multi-dimensional) noise parameter vector
governing the time-dependent PSD $S(f,t;\boldsymbol{\theta})$.
We perform segment-wise inference separately on the pre-gap and post-gap data
to obtain endpoint posterior samples,
\begin{equation}
\left\{\boldsymbol{\theta}^{\rm pre}_i\right\},\qquad
\left\{\boldsymbol{\theta}^{\rm post}_j\right\}.
\end{equation}
From these endpoint posteriors, we construct a uniform ``planning box''
\begin{equation}
\boldsymbol{\theta}\in[\boldsymbol{\theta}_{\rm lo},\boldsymbol{\theta}_{\rm hi}],
\end{equation}
by taking credible summaries for each parameter component
(e.g.\ interval endpoints) and pinning/shared constraints for parameters
that are common across the gap.

To avoid over-constraining the planning prior, we conservatively widen the
endpoint-derived credible intervals by increasing their half-width by 15\%
for each parameter component (i.e.\ a 15\% safety inflation on the box).
This retains the benefit of a tighter prior for reducing the worst-case smoothness
envelope, while maintaining robustness to posterior uncertainty and modeling mismatch.

\subsubsection{Step 1b: Frequency-domain prior whitening}
\paragraph{Prior whitening as log-slope mismatch reduction.}
Given the planning box from Step 1a, we construct a fixed reference spectrum
$S_{\rm ref}(f)$ from the endpoint information and define a whitened dynamic PSD
\begin{equation}
S^{\rm w}(f,t)=\frac{S(f,t)}{S_{\rm ref}(f)/S_0},
\label{eq:pw_def}
\end{equation}
where $S_0$ is a normalization constant.
Then
\begin{align}
\partial_f \ln S^{\rm w}(f,t)
&=
\partial_f \ln S(f,t) - \partial_f \ln S_{\rm ref}(f),\\
s_1^{\rm w}(f,t)
&=
\Delta F\,
\left|
\partial_f \ln S(f,t) - \partial_f \ln S_{\rm ref}(f)
\right|,
\label{eq:s1_whitened}
\end{align}
so PW reduces the log-slope mismatch relative to a reference spectrum,
rather than only mitigating absolute PSD amplitude differences.
This is specifically aligned with the diagonal-validity criterion in
Eq.~\eqref{eq:s1_def_rewrite}, which depends on $\partial_f \ln S$.

\paragraph{Reference construction from endpoint posteriors.}
To avoid inserting an artificial sharp transition into the reference near the chimeric gap,
we build $S_{\rm ref}(f)$ as a smooth log-mixture of representative endpoint PSD shapes,
\begin{align}
\ln S_{\rm ref}(f)
&=
\alpha_{\rm mix}\,\ln S_{\rm pre}(f)
+
(1-\alpha_{\rm mix})\,\ln S_{\rm post}(f),
\label{eq:sref_mix_rewrite}
\end{align}
where $S_{\rm pre}(f)$ and $S_{\rm post}(f)$ are derived from endpoint summaries
and typically $\alpha_{\rm mix}=1/2$.
We then define the fixed whitening factor
\begin{equation}
W(f)=\frac{1}{\sqrt{S_{\rm ref}(f)/S_0}}.
\label{eq:W_def}
\end{equation}

In implementation, the same $W(f)$ is applied consistently to
(i) the data stream, (ii) the waveform templates, and (iii) the PSD used to
construct the WDM diagonal covariance, ensuring statistical coherence of the likelihood.

\subsubsection{How the tight prior reduces $N_f$ (and why PW further helps)}
In the planning stage, $N_f$ is chosen so that the frequency-direction condition is satisfied
in the prior worst-case under PW:
\begin{equation}
s_1^{\rm w}(f,t;\boldsymbol{\theta})
=
\Delta F\,
\left|
\partial_f \ln S(f,t;\boldsymbol{\theta})
-
\partial_f \ln S_{\rm ref}(f)
\right|
\le \varepsilon,
\end{equation}
for all $\boldsymbol{\theta}\in[\boldsymbol{\theta}_{\rm lo},\boldsymbol{\theta}_{\rm hi}]$ and for all
$(f,t)$ in the relevant design time/tile region.

Equivalently, define the whitened log-derivative
\begin{equation}
\partial_f \ln S^{\rm w}(f,t;\boldsymbol{\theta})
\;\equiv\;
\partial_f \ln S(f,t;\boldsymbol{\theta})
-
\partial_f \ln S_{\rm ref}(f),
\end{equation}
and the corresponding worst-case envelope
\begin{equation}
\left|\,\partial_f \ln S^{\rm w}\,\right|_{\max}
\;\equiv\;
\max_{\substack{\boldsymbol{\theta}\in[\boldsymbol{\theta}_{\rm lo},\boldsymbol{\theta}_{\rm hi}]\\
(f,t)\in \mathcal{R}}}
\left|
\partial_f \ln S(f,t;\boldsymbol{\theta})
-
\partial_f \ln S_{\rm ref}(f)
\right|.
\end{equation}
Using $\Delta F=1/(2N_f\Delta t)$, the diagonal-validity requirement yields the continuous
lower bound
\begin{equation}
N_f^{\rm (cont)}
\;\gtrsim\;
\frac{\left|\,\partial_f \ln S^{\rm w}\,\right|_{\max}}{2\,\varepsilon\,\Delta t}.
\end{equation}

\subsubsection{Why choose $N_f=2^k$ in our implementation.}
Although the criterion above is continuous in $N_f$, the discrete WDM time--frequency grid
and transform evaluation are implemented efficiently for dyadic resolutions.
In particular, the WDM pixelization uses $N=N_tN_f$ with frequency bin width $\Delta F$,
and the WDM window support length scales as
\begin{equation}
K = 2qN_f,
\end{equation}
so the computationally dominant transform steps involve FFTs of length proportional to $K$
(e.g., kernel evaluation at cost $\mathcal{O}(K\ln K)$ in the standard fast WDM procedure).
Restricting $N_f$ to dyadic values,
\begin{equation}
N_f=2^k,\qquad k\in\mathbb{N},
\end{equation}
aligns the discrete frequency-layer indexing with an FFT-friendly radix-2 structure,
reduces the need for additional padding to reach an efficient transform length,
and therefore makes the numerical evaluation of the WDM kernels consistent and efficient.
This is an implementation choice; it does not change the underlying
continuous design requirement.

\subsection{Stage 2: adaptive gap expanding}

Even with frequency-domain prior whitening, the diagonal WDM approximation can fail locally
near a gap edge when the time-direction smoothness criterion is violated.
Therefore, we apply a single adaptive gap expanding step driven by the $\mu_1$ test~\cite{Cornish2025}.

\subsubsection{Time-direction smoothness test ($\mu_1$) --- worst-case check}

Let the (planning) noise-parameter vector $\boldsymbol{\theta}$ be constrained to the uniform box
$\boldsymbol{\theta}\in[\boldsymbol{\theta}_{\rm lo},\boldsymbol{\theta}_{\rm hi}]$.
The planning box is constructed from the endpoint posteriors and then conservatively widened by
inflating each parameter-component interval half-width by $15\%$.

For a given (whitened) chimeric PSD realization, we compute $\mu_1$ in WDM pixel space
and assess the conservative acceptance criterion based on the worst-case PSD realization
within the planning prior:
\begin{equation}
\max_{\boldsymbol{\theta}\in[\boldsymbol{\theta}_{\rm lo},\boldsymbol{\theta}_{\rm hi}]}
\;\max_{m}\left|\mu_1(\boldsymbol{\theta})\right|
\;\le\;\varepsilon,
\label{eq:mu1_test_worst}
\end{equation}
where $\varepsilon$ is the same user-chosen threshold used in planning.

If Eq.~\eqref{eq:mu1_test_worst} holds, we keep the full WDM likelihood
with the diagonal covariance in the whitened space.

\subsubsection{Boundary gating via minimal chimera window width $L_{\min}$}

If Eq.~\eqref{eq:mu1_test_worst} fails, the diagonal approximation is unreliable only within
a neighborhood around the chimeric transition near the gap edges.
We restore local diagonal validity by expanding the unobserved neighborhood
as little as possible through AGE.

\paragraph{Minimal total window from the $\mu_1$ criterion.}
Let $L$ denote the total width of the time-domain chimera transition window
in WDM time-pixel units.
For the (planning) worst-case PSD realization within the prior,
we compute the discrete smoothness measure $\mu_1(L)$ and define
the minimal required width $L_{\min}$ by the conservative condition
\begin{equation}
\max_{\boldsymbol{\theta}\in[\boldsymbol{\theta}_{\rm lo},\boldsymbol{\theta}_{\rm hi}]}
\mu_1(L_{\min};\boldsymbol{\theta})=\varepsilon.
\end{equation}
By construction, $L_{\min}$ is never smaller than the natural gap width:
when the natural gap already satisfies the $\mu_1$ criterion, $L_{\min}$ equals the natural width
(and the expansion becomes inactive).
Otherwise, the AGE step sets the expanded chimera window width to $L_{\min}$.

\subsubsection{Expanded missing mask and Bayesian augmentation}

Let $K$ be the time-support length of the Meyer wavelet filter (so that wavelet support
extends by $K/2$ beyond the chimera transition window).
Let $\mathrm{gap}_R$ denote the right edge (in WDM time-pixel coordinates) of the natural gap.
Given the minimal total chimera window width $L_{\min}$ determined by the $\mu_1$ criterion,
the expanded chimera window is chosen in a left-biased manner as
$[\mathrm{gap}_R-L_{\min},\,\mathrm{gap}_R]$.
In this stage, we expand the missing-data neighborhood so that this chimera window becomes unobserved,
and we further exclude all WDM wavelet coefficients whose time supports overlap it.
Therefore, the expanded unobserved region is
\begin{equation}
d_{m'}=
\text{missing data on }
\Bigl[
\mathrm{gap}_R-L_{\min}-\frac{K}{2},\;
\mathrm{gap}_R+\frac{K}{2}
\Bigr].
\end{equation}

We then perform Bayesian augmentation (Pearson--Cornish gap filling)
in the time domain at the current MCMC draw $i$:
\begin{equation}
d_{m'} \sim p\!\left(d_{m'}\mid d_o,\boldsymbol{\theta}^{(i)}\right),
\end{equation}
using the localized neighborhood covariance implied by the WDM kernel support.
Importantly, the PW from Stage~1 is applied consistently:
the whitening PSD and the diagonal  WDM covariance are evaluated in the same
prior-whitened representation.
The expansion direction (left-biased, right-biased, or symmetric) can be chosen freely according
to the signal-dependent information density profile in practice; the above left-biased choice
is a convenient default for the chirp model considered here.

\section{Testing on Simulated Data}
\label{sec:test}
\subsection{Toy Model}
\label{sec:toy_model}

To isolate and stress-test the short-gap + large PSD-jump failure mode,
we adopt a toy model that preserves the same structural ingredients as our injections:
(i) a power-law PSD with a chimeric amplitude transition across the physical gap,
and (ii) a linear chirp signal.
Unless otherwise stated, the PSD shape parameters are shared across the gap,
and only the overall PSD amplitude changes.

\paragraph{Noise model: chimeric amplitude jump on a power-law PSD.}
We model the one-sided PSD as a power law in frequency with knee parameter $s$
and spectral index $\alpha$:
\begin{equation}
S(f,t)=A(t)\left(f^2+s^2\right)^{-\alpha/2}.
\end{equation}
Across the physical gap, the PSD amplitude transitions between two asymptotic values
$A_{\rm pre}$ and $A_{\rm post}$ using an optimally smooth three-point Hermite connector.
Let the transition window be $t\in[t_s,t_e]$ and define the unit variable
\begin{equation}
u=\frac{t-t_s}{t_e-t_s}\in[0,1].
\end{equation}
Then
\begin{equation}
A(t)=
\begin{cases}
A_{\rm pre}, & t<t_s,\\[3pt]
A_{\rm pre}\bigl(1-\phi(u)\bigr)+A_{\rm post}\phi(u),
& t\in[t_s,t_e],\\[3pt]
A_{\rm post}, & t>t_e,
\end{cases}
\end{equation}
with
\begin{equation}
\phi(u)=3u^2-2u^3.
\end{equation}
In the default injections,
$A_{\rm pre}=1.5$, $A_{\rm post}=3.0$, $s=10^{-3}\,{\rm Hz}$, and $\alpha=2.0$,
so that the amplitude jump is
$A_{\rm post}/A_{\rm pre}=2$.
Unless stated otherwise, the PSD shape $(s,\alpha)$ is shared across the gap.

\paragraph{Hermite connector from endpoint value and slope continuity.}
Inside the transition window we choose a cubic polynomial
\begin{equation}
\phi(u)=au^3+bu^2+cu+d,
\end{equation}
such that both endpoint values and first derivatives match continuously:
\begin{align}
\phi(0)&=0, & \phi(1)&=1,\\
\phi'(0)&=0, & \phi'(1)&=0.
\end{align}
With $\phi'(u)=3au^2+2bu+c$, the constraints imply
$d=0$, $c=0$, $a+b=1$, and $3a+2b=0$,
leading to $a=-2$ and $b=3$.
Hence
\begin{equation}
   \phi(u)=3u^2-2u^3,  
\end{equation}
which is the unique cubic connector satisfying the
four Hermite constraints.

\begin{figure*}[t]
\centering
\includegraphics[width=\textwidth]{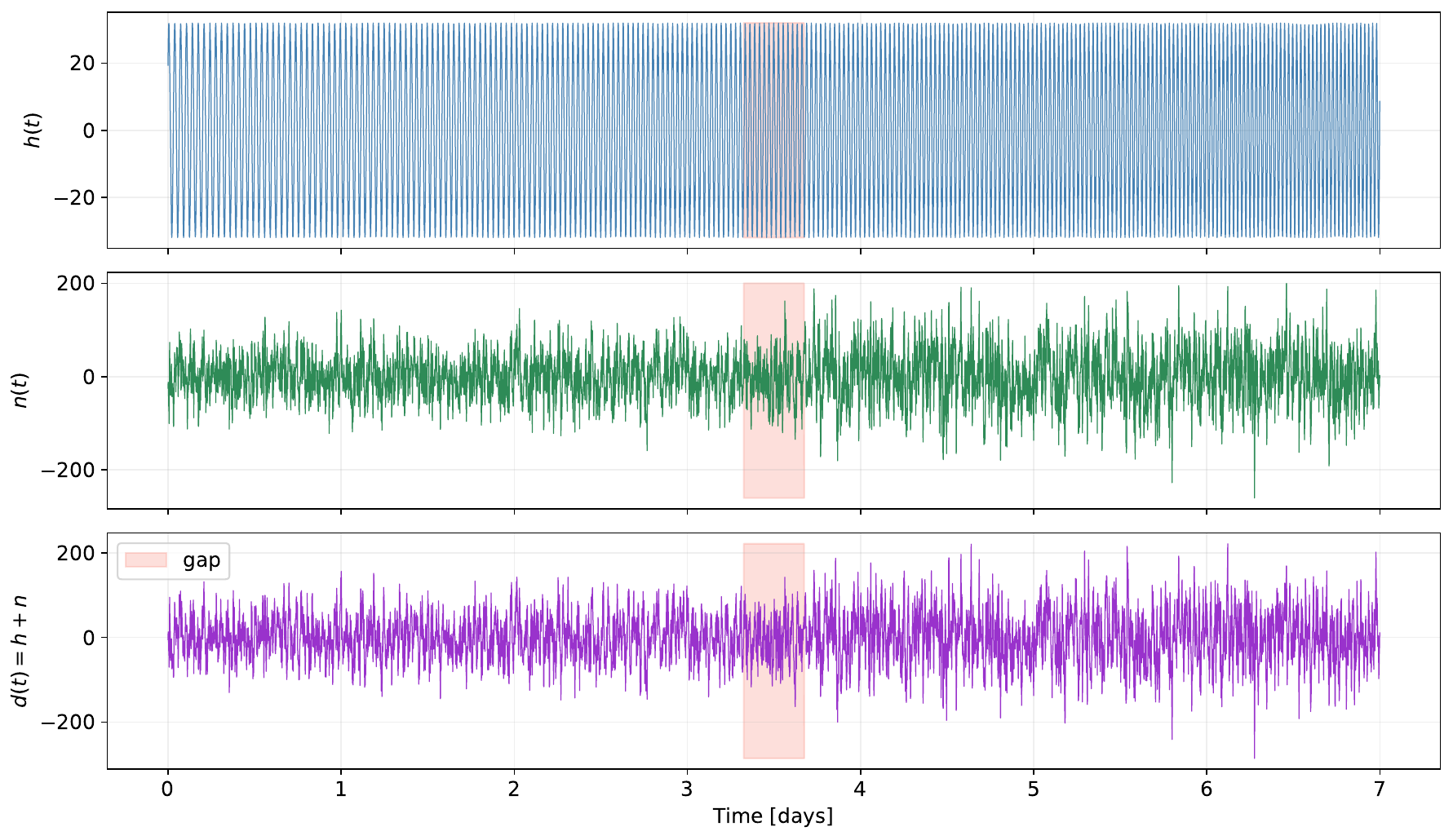}
\caption{Example toy-model simulated data illustrating a short physical gap.
The shaded region marks the masked gap; the top panel shows $h(t)$, the middle panel shows the noise $n(t)$, and the bottom panel shows the combined data $d(t)=h(t)+n(t)$ (and its time-domain variation).}
\label{fig:toy_gap_replay}
\end{figure*}

\paragraph{Gap model and discretization.}
We simulate an observation of duration $T_{\rm obs}=7$ days with
a fixed number of time samples $N$ (default $N=5120$), so that
\begin{equation}
\Delta t=\frac{T_{\rm obs}}{N}.
\end{equation}
The physical gap is centered in the time series with length of:
\begin{equation}
\Delta N_{\rm gap} = 5\%~N.
\end{equation}
In the inference stage, the gap samples are treated as missing
(removed or masked), producing the short-gap condition.

An example of the resulting gapped, non-stationary toy data (including the shaded physical gap and the corresponding time-domain variation) is shown in
Fig.~\ref{fig:toy_gap_replay}.

\paragraph{Signal model: linear chirp injection.}
Injected signals are single-component linear-chirp sinusoids, designed to mimic a
gravitational-wave-like signal whose instantaneous frequency increases over the
observation. The MCMC parameter vector is
\begin{equation}
\theta_s=( A_s,\phi_s,\omega_s,\gamma_s),
\end{equation}
where \(\omega_s\) is the angular frequency at \(t=0\).
Let \(T=t_{\rm span}\) denote the full observation duration.
The injected waveform is
\begin{equation}
h(t)=A_s\sin\!\left[
\phi_s+\omega_s t+\frac{\omega_s\,\gamma_s\, t^2}{2T}
\right],
\end{equation}
which corresponds to the instantaneous angular frequency
\begin{equation}
\omega(t)=\omega_s\left(
1+\gamma_s\,\frac{t}{T}
\right).
\end{equation}
Thus \(\gamma_s\) is the fractional frequency change over \(T\),
\(\Delta f/f(0)=\gamma_s\).

The amplitude \(A_s\) is rescaled so that the matched-filter SNR of the
pre-gap segment is approximately \(15\); the same \(A_s\) is used on both
sides of the gap. The corresponding ground-truth (corner-plot) parameters are
\begin{align}
A_s &\simeq 31.988,\\
\phi_s &= 0.65,\\
\omega_s &\simeq 0.002128~{\rm rad/s},\\
\gamma_s &= 0.5.
\end{align}

\subsection{Results}
\label{sec:results}

We compare three analyses for the toy ``short-gap + large PSD jump'' case with
$A_{\rm pre}=1.5$, $A_{\rm post}=3$, and gap fraction $5\%$:
(i) single-segment matched-filter MCMC on the pre-gap data,
(ii) single-segment matched-filter MCMC on the post-gap data,
and (iii) the proposed WDM+PW+AGE analysis.

\begin{figure*}[t]
\centering
\includegraphics[width=\textwidth]{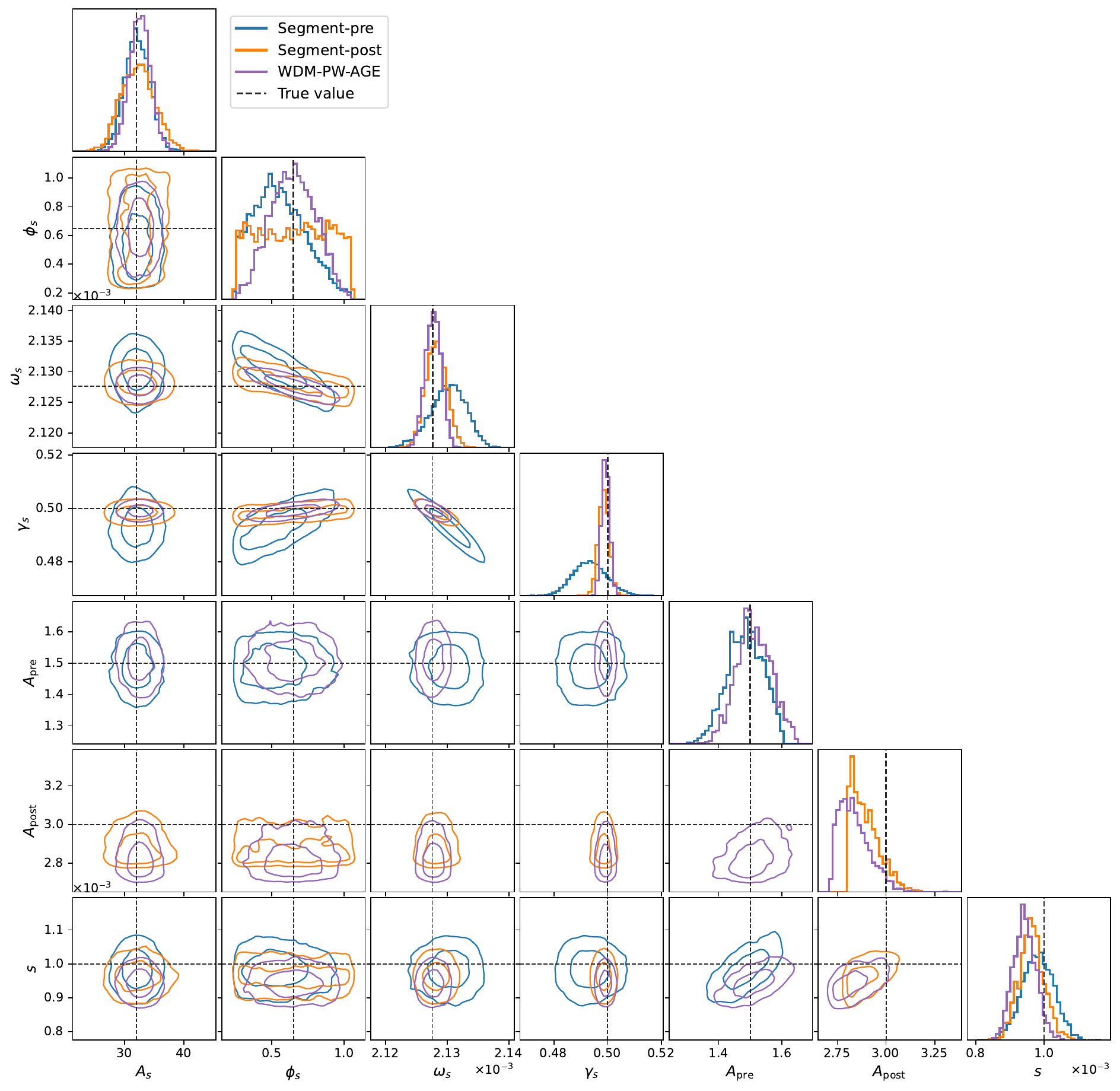}
\caption{Corner plot of the inferred parameters for a fixed injection.
Pre-gap single-segment MCMC (blue) and post-gap single-segment MCMC (orange) are compared
against WDM-PW-AGE (purple). Dashed lines indicate the true parameter values.
For the signal parameters (estimated using information from both sides of the gap), the
WDM-PW-AGE posterior exhibits a tighter credible region than the single-segment analyses.}
\label{fig:results_corner}
\end{figure*}

The WDM run fixes $\alpha=2$ during MCMC sampling; only $(A_{\rm pre},A_{\rm post},s)$ are explored.

\paragraph{PW effect.}
The WDM prior is constructed from the endpoint posteriors of the two segment runs, then expanded by $15\%$ on each side.
Using this narrower prior box, PW reduces the frequency-direction mismatch budget, lowering the required number of WDM frequency layers:
\begin{equation}
N_f: 64 \to 32,
\qquad
N_t: 80 \to 160.
\end{equation}
This halves the frequency resolution while doubling the time-direction budget.

\paragraph{AGE-triggered gap widening.}
After prewhitening, the time-direction validity test yields $\mu_1=0.1688 > \varepsilon=0.1$,
triggering adaptive gap expansion. The physical gap is widened from the natural gap
(which corresponds to an integer WDM time-pixel span of 7) to an expanded chimera transition
window with continuous width
$L_{\rm final}=11.9237\,\Delta T$ (in WDM time-pixel units), i.e.\ a window-width increment of
$+4.924\,\Delta T$.

\paragraph{Posterior comparison.}
Fig.~\ref{fig:results_corner} shows a corner plot comparing inferred parameters from pre-gap (blue) 
and post-gap (orange) single-segment MCMC analyses against the WDM+PW+AGE method (purple) 
for a fixed injection (dashed lines denote true values). 
Table~\ref{tab:post_inj_pre_post_wdm} complements this comparison by listing the corresponding posterior summaries (reported under the 95\% credible interval) for the same injection.
By contrast, the WDM+PW+AGE posterior (purple) is substantially more concentrated and better aligned with the injected values, with tighter and more symmetric joint parameter correlations than the pre- and post-gap baselines.

\begin{table*}[t]
\centering
\small
\renewcommand{\arraystretch}{1.45}
\setlength{\tabcolsep}{6pt}
\resizebox{\textwidth}{!}{%
\begin{tabular}{lccc}
\hline
\textbf{Parameter (true value)} & \textbf{pre posterior} & \textbf{post posterior} & \textbf{WDM-PW-AGE posterior} \\
\hline
$A_s$ ($31.9882$) &
$32.0345^{+4.2278}_{-4.2116}$ &
$32.2362^{+5.6607}_{-5.1397}$ &
$32.5965^{+3.5535}_{-3.6555}$ \\
$\phi_s$ ($0.6500$) &
$0.5409^{+0.3832}_{-0.2680}$ &
$0.6567^{+0.3710}_{-0.3889}$ &
$0.6456^{+0.2913}_{-0.2904}$ \\
$\omega_s$ ($2.1276\times10^{-3}$) &
$\left(2.1300\times10^{-3}\right)^{+5.0000\times10^{-6}}_{-6.0000\times10^{-6}}$ &
$\left(2.1280\times10^{-3}\right)^{+3.0000\times10^{-6}}_{-3.0000\times10^{-6}}$ &
$\left(2.1280\times10^{-3}\right)^{+2.0000\times10^{-6}}_{-3.0000\times10^{-6}}$ \\
$\gamma_s$ ($0.5000$) &
$0.4932^{+0.0130}_{-0.0113}$ &
$0.4985^{+0.0038}_{-0.0039}$ &
$0.4992^{+0.0029}_{-0.0027}$ \\
\hline
$A_{\mathrm{pre}}$ ($1.5000$) &
$1.4847^{+0.0997}_{-0.1153}$ &
-- &
$1.5092^{+0.1107}_{-0.1031}$ \\
$A_{\mathrm{post}}$ ($3.0000$) &
-- &
$2.8813^{+0.1879}_{-0.0773}$ &
$2.8278^{+0.1909}_{-0.0988}$ \\
$s$ ($1.0\times10^{-3}$) &
$\left(0.9830\times10^{-3}\right)^{+0.8900\times10^{-4}}_{-0.9200\times10^{-4}}$ &
$\left(0.9630\times10^{-3}\right)^{+0.7600\times10^{-4}}_{-0.7200\times10^{-4}}$ &
$\left(0.9420\times10^{-3}\right)^{+0.6800\times10^{-4}}_{-0.6000\times10^{-4}}$ \\
\hline
\end{tabular}%
}
\caption{Posterior summaries reported under the 95\% credible interval. Note that $\alpha$ is omitted because it is effectively fixed at $\alpha = 2.0$. For the signal parameters, the WDM-PW-AGE posterior is tighter and closer to the injected values because it coherently combines information from both the pre- and post-gap data. 
This improvement does not rely on adding extra noise-model information beyond the assumed chimeric noise transition; rather, it reflects more effective utilization of the available data.}
\label{tab:post_inj_pre_post_wdm}
\end{table*}

\section{CONCLUSION}
\label{sec:conclusion}

This paper addresses gravitational-wave inference in the presence of non-stationary noise and data gaps, focusing on a practically challenging regime of short gaps accompanied by large PSD amplitude jumps. In this regime, the diagonal (Whittle-like) likelihood in the WDM basis—whose validity is controlled by the smoothness criteria introduced by Cornish—can fail locally near gap edges, leading to biased and/or less robust parameter recovery when one relies on single-sided analyses.

Building on the WDM-based gap augmentation framework of Pearson and Cornish, we propose a two-stage remedy that targets this specific failure mode rather than resorting to full segmentation. First, we introduce frequency-domain prior-predictive prewhitening, where a fixed whitening reference spectrum is constructed from endpoint information. PW reduces the effective frequency-direction log-slope mismatch, enabling a coarser choice of the WDM frequency-layer resolution while retaining the diagonal approximation as much as possible. Second, we apply a dynamic smoothness assessment using the $\mu_1$ criterion and then perform selective boundary gating via adaptive gap expanding only when needed. When the local diagonal-validity test fails, AGE expands the unobserved region with a minimal guard-band to restore local validity, thereby aiming to preserve more usable data than more aggressive gap-splitting strategies.

In our numerical experiments with chimeric noise realizations and signal injections within the WDM framework, we compare three analyses: pre-gap single-sided inference, post-gap single-sided inference, and the proposed WDM+PW+AGE method. The resulting posteriors from WDM+PW+AGE are substantially more concentrated and closer to the injected parameter values than the pre- and post-gap baselines, indicating that the combination of prior-informed preconditioning and validity-driven local gating can mitigate the gap-induced diagonal-approximation failure near the boundary.

While the method is designed to minimize information loss, it is not a universally perfect substitute for more exact likelihood treatments. In particular, AGE restores local diagonal validity by enlarging the masked neighborhood around the gap edges, which necessarily removes some information from that region. Consequently, if the signal information content is strongly concentrated within (or close to) the gap neighborhood, or if diagonal validity is frequently violated in a way that requires repeated/large expansions, the net benefit may diminish. Assessing such cases would require additional study beyond the toy-model stress tests considered here.

Overall, our results suggest that prior-informed preconditioning together with data-driven diagonal-validity testing offers a principled and computationally efficient route for handling short gaps in non-stationary noise for next-generation detector analyses, where both PSD discontinuities and localized gap expansions are expected to arise.

\section{ACKNOWLEDGMENTS}
This work was supported by the China Postdoctoral Science Foundation Grant No.~2026M793675.

\end{document}